\documentclass{jfm}
\usepackage{graphicx, placeins}
\usepackage{epstopdf, epsfig}
\usepackage{siunitx, amsmath}
\usepackage{caption, subcaption}
\usepackage{hyperref}

\shorttitle{Effect of Roughness on Swimming Plates}
\shortauthor{J. M. O. Massey, B. Ganapathisubramani, G. D. Weymouth}

\title{A systematic investigation into the effect of roughness on self-propelled swimming plates}

\author{J. M. O. Massey\aff{1,2}
  \corresp{\email{masseyjmo@gmail.com}},
  B. Ganapathisubramani\aff{1} \newline
  \and G. D. Weymouth\aff{1,2}}

\affiliation{\aff{1}Faculty of Engineering and Physical Sciences, University of Southampton, UK
\aff{2}Faculty of Mechanical, Maritime, and Materials Engineering, TU Delft, NL}

\begin{document}

\maketitle

\begin{abstract}
    This study examines the effects of surface topography on the flow and performance of a Self-Propelled Swimming (SPS) body. We consider a thin flat plate with an egg-carton roughness texture undergoing prescribed undulatory swimming kinematics at a Strouhal number of $0.3$ and tail amplitude to length ratio of $0.1$; we use plate Reynolds numbers of $\Rey=6,12,$ and $24\times 10^3$, and focus on $12,000$. As the roughness wavelength is decreased, we find that the undulation wave speed must be increased to overcome the additional drag from the roughness and maintain SPS.
    Correspondingly, the extra wave speed raises the power required to maintain SPS, making the swimmer less efficient.
    To decouple the roughness and the kinematics, we compare the rough plates to equivalent smooth cases by matching the kinematic conditions. We find that all but the longest roughness wavelengths reduce the required swimming power and the unsteady amplitude of the forces when compared to a smooth plate undergoing identical kinematics.
    Additionally, roughness can enhance flow enstrophy by up to 116\% compared to the smooth cases without a corresponding spike in forces; this suggests that the increased mixing is not due to increased vorticity production at the wall. Instead, the enstrophy is found to peak strongly when the roughness wavelength is approximately twice the boundary layer thickness over the $\Rey$ range, indicating the roughness induces large-scale secondary flow structures that extend to the edge of the boundary layer.
    This study reveals the nonlinear interaction between roughness and kinematics beyond a simple increase or decrease in drag, illustrating that roughness studies on static shapes do not transfer directly to unsteady swimmers.

\end{abstract}


\section{Introduction}

    Underwater propulsion is an area of research that brings together engineers and biologists alike; it has fostered a deeper understanding of aquatic locomotion and inspired innovation in underwater systems. The kinematics have the most significant effect on the locomotive properties of marine animals \citep{Lighthill1960NoteFish, Lighthill1971,Triantafyllou1991,Triantafyllou1993a,Borazjani2008NumericalRegimes,Eloy2012,Saadat2017,DiSanto2021}; however, some intricacies of animal evolution, such as the skins of sharks and odontocetes, have sparked research into the benefits of aquatic surface textures. Understanding the fluid dynamic interaction between kinematics and surface textures will help us to elucidate the contribution of surface textures to aquatic locomotion.
    
    Previous work has identified important non-dimensional kinematic parameters for efficient swimming. The Reynolds number ($\Rey$) affects the efficiency of swimmers which can be in the viscous ($\Rey \approx 10^2$), transitional ($\Rey \approx 10^3$), and inertial ($\Rey \to \infty$) regime \citep{Borazjani2008NumericalRegimes, Borazjani2010OnSwimming}. In this study we focus on the transitional regime with $\Rey=[6, 24]\times 10^3$.
    The Strouhal number ($St$) describes the ratio between the product of the wake width and the shedding frequency, and the flow velocity; \cite{Triantafyllou1991,Triantafyllou1993a} found that the optimal Strouhal number should be in the range $0.25-0.35$. \citet{Eloy2012} tested the kinematics of fifty-three different types of fish based on Lighthill's elongated body theory \citep{Lighthill1960NoteFish, Lighthill1971} and found that for thin tails, the optimal Strouhal number range was $0.2-0.4$. \cite{Saadat2017} showed that the Strouhal number was insufficient for efficient locomotion and defined a range of optimum motion amplitude to length ratio of $0.05$ to $0.15$. \cite{DiSanto2021} recently compared forty-four species of fishes and found that despite fishes' different morphologies\textemdash categorised as; anguilliform, subcarangiform, carangiform, and thunniform\textemdash they shared a statistically significant oscillation amplitude, a.k.a. kinematic envelope. This work is noteworthy as previous work suggested the presence of different kinematics was dependent on the morphology.

    While kinematics have received the majority of the literature's attention, they are not the only factor affecting the locomotive properties of swimmers. Many studies have identified drag-reducing properties of certain textures. Although most of these studies are applied on static geometries, it has been suggested that these effects could transfer to unsteady aquatic propulsion. \cite{Bechert2000} showed that when shark skin-like denticles interlock, they can passively reduce the drag on the surface much like a riblet. Riblets are small-scale two-dimensional transverse grooves whose height scale with the viscous scales of the flow and can act as drag reduction devices (\cite{WALSH1982, Park1994FlowLayer, RGarcia-Mayoral2011, Cui2019EffectLayer}). However, if riblets are not viscous scaled and/or have other larger-scale features to them (such as a larger pattern of inclinations) they create new flow features and perhaps increase drag (\cite{Bechert2000, Nugroho2013Large-scaleRoughness, Boomsma2016, VonDeyn2022FromRidges, Rouhi2022Riblet-generatedAnalogy}). The sensitivity of riblets to these specific flow conditions calls into question their applicability to boost the performance of unsteady swimmers.
    

    During a swim cycle, the surface of a swimmer might briefly encounter the viscous scales set out by \cite{Bechert2000} for drag-reduction of riblets however, the morphing of the body and the fluctuating viscous length scales force the near-wall flow outside the conditions that cause riblets to decrease drag the majority of the time. To comprehend the potential hydrodynamic benefit of surface textures, we must look at them in the context of a larger, dynamic system. For example, vortex generators are often low-profile roughness elements that can significantly effect the flow and stimulate massive increases of aerodynamic performance when positioned correctly \citep{Lin1994, Lin2002}. Similarly, studies suggest that shark skin uses passive control in the flank region to bristle the skin while swimming, increasing boundary layer mixing and helping to keep the flow attached at areas of flow reversal (\citealt{lang2008bristled, afroz2016experimental, Santos2021}). In this vein, \cite{Oeffner2012} tested samples of skin from the midsection of a short-fin mako shark on both a rigid flapping plate and a flexible plate. They found that the skin actually reduced the rigid plate's propulsive effectiveness. They also found that adding shark skin increased the flexible plate's swimming speed by $12\%$. However, they do not provide the amplitude envelope for the flexible plate which would ensure constant kinematics between the cases tested. Similarly, \cite{Wen2014} covered an undulating plate in 3D printed denticles of $100\times$ actual size and measured a $6.6\%$ efficiency increase. Again, they do not provide an amplitude envelope to ensure the kinematics between the smooth and the rough surfaces remain constant. The denticles in the above-mentioned work (\cite{Oeffner2012, Wen2014}) are not scaled with local viscous scales so that they are within the drag-reducing regimes set out by \cite{Bechert2000}. Therefore, the physical mechanisms responsible for the differences between rough and smooth surfaces remain unclear and illustrates the need for a systematic study to examine the interaction between surface textures and kinematics.
    
    The complex and multiscale shape of denticles does not lend itself to systematic investigations into the interplay between surface textures and kinematics. Consequently, we look to simplify the surface texture and focus on the first mode effects of roughness. We need to span a relevant physical space and yet ensure that the parameterisation of the surface is suited for the proposed problem. We look for inspiration in the surface textures/geometries that have been explored in previous studies that have focused on developing methods for predicting drag on flow over rough surfaces (\cite{Moody1944FrictionFlow, Jimenez2004, Flack2010ReviewRegime, Flack2014RoughnessFlowsa, Garcia-Mayoral2019, Chung2021PredictingSurfaces}). Previous studies have indicated that the ratio of the total projected frontal roughness area to the wall-paralleled projected area (solidity, $\Lambda$, \citealt{Schlichting1936}) and the mean slope of the roughness texture (in the streamwise and spanwise-directions, also known as effective slope, $ES$, \citealt{Napoli2008TheFlows}) are two geometric parameters of a rough surface known to significantly affect the flow and forces. These two parameters can be easily altered for structured surfaces where the surface geometry has a sinusoidal shape. In fact, previous works have used sinusoidal roughness where the variation of the two roughness properties can be achieved by only altering the wavelength of the sinusoidal shape (\citealt{Napoli2008TheFlows, Chan2015, Ma2020ScalingSteepness, Ganju2022AmplitudeFixed}). This presents us with a surface that can be used to understand how the primary scales of roughness (parameterised by a single quantity) interacts with kinematics, with the hope that the findings can be generalised to more complex geometries. 
    
    In this work, we study the interaction between kinematics and roughness topologies through high-resolution simulations of a rough self-propelled swimming thin plate. We consider three Reynolds numbers ($\Rey=6,12,$ and $24\times 10^3$) with a focus on $\Rey=12,000$ (based on swimming speed and chord length) to access moderate Reynolds numbers for the types of flows that are in line with previous efforts (\cite{Oeffner2012, Wen2014, Saadat2017, Domel2018, Thekkethil2018}). We also fix the kinematics of the plate to a simple travelling waveform with a fixed Strouhal number, $St$, that is in the propulsive regime for flapping foils and has been used extensively in previous studies (\cite{Dong2007CharacteristicsArrangement, Borazjani2008NumericalRegimes, Maertens2017a, Muscutt2017PerformanceFoils, Thekkethil2018, Zurman-Nasution2020}). As denticle geometries are complex with several potentially important length scales, we focus on a simple roughness texture with a single-length scale to assess how the topology interacts with the kinematics and impacts the hydrodynamic properties of the swimmer. By combining information from two different, but, well-established topics we hope to understand the influence of one on the other. These dynamic simulations with roughness elements are the first of their kind, allowing us to establish a link between surface roughness and kinematics, and then\textemdash with comparison to a smooth kinematic counterpart\textemdash directly isolate the nonlinear interaction of the roughness and kinematics.

\section{Methodology}\label{sec:methodology}
    \subsection{Geometry}

        We use a flat plate with a thickness $3\%$ of the plate length $L$ as the base model. Using a flat plate couples the dynamics and the surface texture in the simplest possible setup, similar to thin aerofoil theory. Using a body with curvature would also make some bumps more proud to the flow than others, changing the effective amplitude of the bumps along the body. A thin plate enables the use of a constant bump amplitude $h$ of $1\%$ of the plate length all along the body.
        
\begin{figure}
            \centerline{\includegraphics[width=\linewidth]{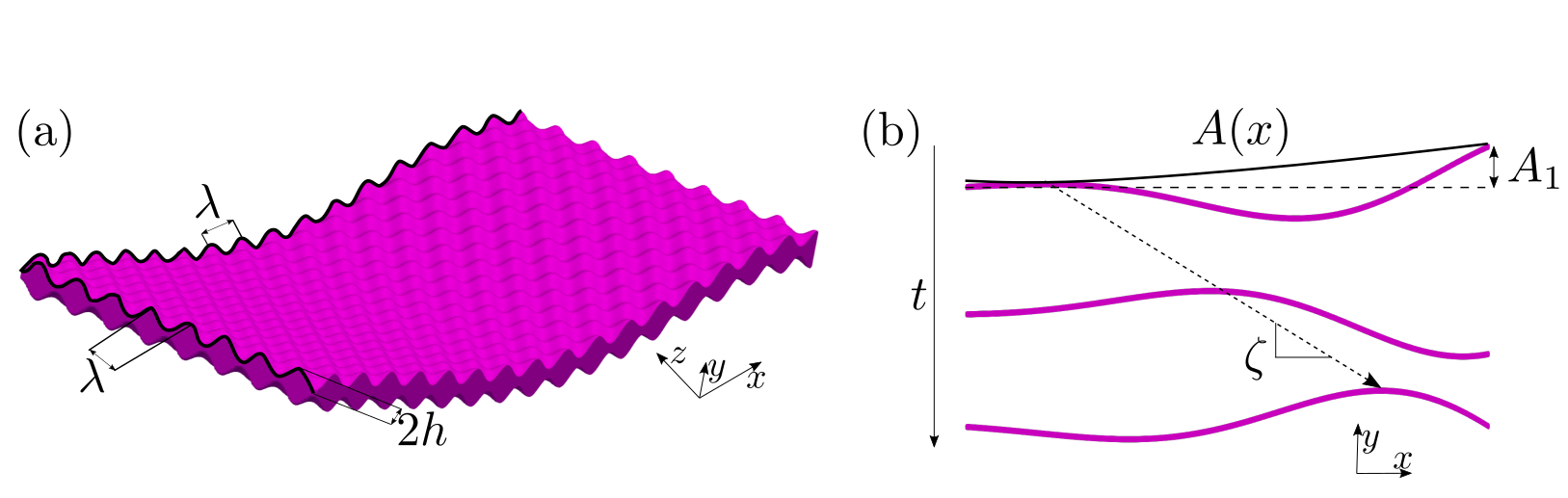}}
            \caption{(a) The geometry used which has two defining parameters: $\lambda$, the wavelength of the roughness, and $h$, the roughness amplitude. (b) A visual representation of the parameters that define the plate motion.}
        \label{fig:geometry}
        \end{figure}

        For the roughness, we use a sinusoidal roughness, similar to \cite{Napoli2008TheFlows, Chan2015, Ma2020ScalingSteepness, Ganju2022AmplitudeFixed}, that allows us to vary the roughness topology systematically. Figure \ref{fig:geometry}a illustrates the parameters affecting the roughness topology. Normalising all lengths by the plate length $L$, the topography is defined as

        \begin{equation}
            y(x,z)= \begin{cases}
            h \sin\Big(\frac{2\pi x}{\lambda}\Big)\cos\Big(\frac{2\pi z}{\lambda}\Big), &  \text{for } y \geq 0.015-h \\
            h \sin\Big(\frac{2\pi x}{\lambda}-\pi\Big)\cos\Big(\frac{2\pi z}{\lambda}\Big), & \text{for  } y \leq -0.015+h
                    
            \end{cases}
            \label{eq: roughness definition}
        \end{equation}
        \\
        where $y$ is the direction normal to the plate and $x,z$ are the tangential direction, and $\lambda$ is the roughness wavelength.
        
    \subsection{Kinematics}\label{sec:Kinematics}

        Continuing to scale all lengths by $L$, using the swimming speed $U$ to scale velocity and $L/U$ to scale time, we define the excursion of the body from the centre line as

        	\begin{equation}
        	    y(x) = A(x) \sin \big(2\pi  \, [f t - x/\zeta] \big)
            \label{eq: swimming}
        	\end{equation}
        	\\
        where $f$ is the frequency, $\zeta$ is the phase speed of the travelling wave, and $A(x)$ is the amplitude envelope, all of which are illustrated in figure \ref{fig:geometry}b. The Strouhal number is set to peak propulsive value $St=0.3$ which determines the scaled frequency as $f=St/2A_1$, where $A_1=A(x=1)$ is the trailing edge amplitude and defines the wake width of the system. We modify the recent result from \citealt{DiSanto2021} for the envelope
        
        \begin{equation}\label{eq:kinematic trajectory}
            A(x)=\frac{A_1(a_2x^2+a_1x+a_0)}{\sum_{i=0}^{2} a_i}
        \end{equation}
        \\
        using $a_{0,1,2}=(0.05, 0.13, 0.28)$ and $A_1=0.1$, as found to be optimal in \citealt{Saadat2017}. The modification changes $a_1$ from $-0.13$ \citep{DiSanto2021} to $0.13$ reducing the amplitude of the leading edge, enabling self-propelled swimming over a wider range of $\lambda$.

    \subsection{Numerical Method}

      \begin{figure}
          \centerline{\includegraphics[width=\linewidth]{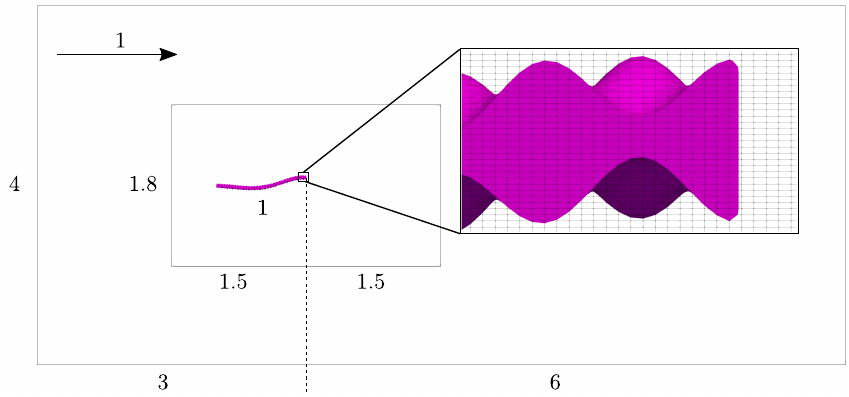}}
          \caption{Schematic for the domain and grid. The inner box shows the region where the grid is uniformly rectilinear; from there, the grid stretches toward the domain extent. This is a representative grid for the $xy$ plane on the geometry where $\lambda=1/16$. We use a periodic boundary condition and $\max(6\lambda, 0.25)$ to define the repeating spanwise domain size. The insert indicates the grid around the tail of the plate showing that, for $\lambda=1/16$, we have $16$ cells resolving the roughness wavelength, and $5$ cells resolving the amplitude of the surface.} 
        \label{fig:grid}
        \end{figure}

        We simulate incompressible fluid flow with the dimensionless Navier-Stokes equation combined with the continuity equation
        
        \begin{align}
            \frac{\partial \vec{u}}{\partial t} + (\vec{u} \cdot \vec{\nabla})\vec{u} = &
            -\vec{\nabla}p + \frac 1\Rey \nabla^2 \vec{u} \\
            \vec{\nabla} \cdot \vec{u} = & 0 \text{ ,}
        \end{align}
        \\
        where $\vec{u}(\vec{x}, t) = (u, v, w)$ is the scaled velocity of the flow, and $p(\vec{x},t)$ is the scaled pressure. We use Lotus, an in-house developed finite-volume implicit large eddy simulation (iLES) code. The implicit modelling of iLES derives from a flux-limited QUICK (quadratic upstream interpolation for convective kinematics) treatment of the convective terms \citep{Hendrickson2019WakeEntrainment}. We use an adaptive time-step based on the Courant Friedrichs Lewy condition that has been shown to converge with $O(2)$ \citep{Lauber2022}.
        
        The undulating plate was coupled to these flow equations using the Boundary Data Immersion Method (BDIM) formulated in \cite{Weymouth2011} and further developed for higher Reynolds numbers in \cite{Maertens2015} and thin geometries in \cite{Lauber2022}. BDIM enforces the boundary condition on the body by convolving together the fluid and body governing equations on a Cartesian background grid. BDIM has been extensively validated in those studies and converges at second order in both time and space. The boundary conditions we enforce on the body is the no-slip condition. Symmetry conditions are enforced on the upper and lower domain extents and there is a periodic condition in the spanwise-direction.
      
        Figure \ref{fig:grid} details the grid and domain set-up. We used the smallest domain for which the forces on the body remained invariant when the size increased. We define the force and power coefficients as

        \begin{equation}
            C_T = \frac{\oint \vec{f_x} ds}{0.5 S},\quad C_L = \frac{\oint \vec{f_y} ds}{0.5 S},\quad C_P = \frac{\oint \vec{f}\cdot\vec{v} ds}{0.5 S}
        \end{equation} \label{eq: forces}
        \\
        where $\vec{f}=-p\hat n$ is the normal pressure stress on the body, $\vec{v}$ is the body velocity, and $S$ is the planform area of the smooth plate. We use a rectilinear-grid over the domain $x\in[-2, 7]$, $y\in [-2,2]$ and vary the spanwise-direction so that $z\in[0, \max(6\lambda, 0.25)]$ (figure \ref{fig:grid}). For the area that contains the body motion and the immediate wake, we use a uniform grid and then implement hyperbolic stretching of the grid cells away from this area (figure \ref{fig:grid}). The grid is refined in $y$ such that $\Delta y$ is half $\Delta x$ and $\Delta z$ (figure \ref{fig:grid} insert). This gives us a total number of grid points of ranging $[67.1, 403]\times 10^6$; which are distributed as $(n_x,n_y,n_z)=(1536,1536,[64, 384])$. Careful consideration of the aspect ratio of the maximum stretched cell meant that it did not exceed five times that of the uniform region to avoid distorting the flow in the wake.
        
        Further details of the numerical method verification and validation are given in the appendices. We present the grid convergence results in appendix \ref{sec:convergence} and details of the domain-size convergence in appendix \ref{sec:domain}. The solver is then validated against experimental oscillating plate results of \citet{Lucas2015EffectsModel}, reproducing the thrust force to within 1.7\%, in appendix \ref{sec:kin-validation}.
 
\section{Results}        
     
    \subsection{Self-Propelled Swimming}

        \begin{figure}
          \centerline{\includegraphics{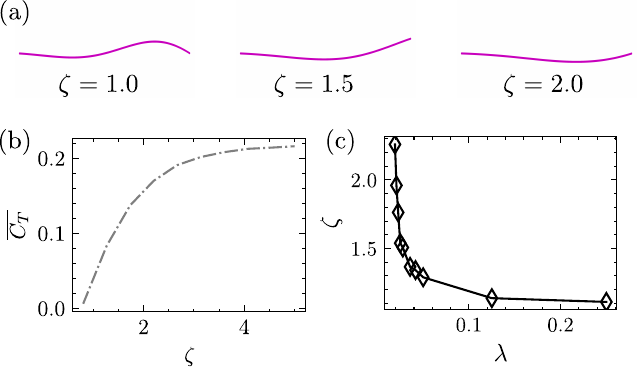}}
          \caption{The impact of $\zeta$ on (a) the shape of and (b) the forces on a smooth plate. (c) The required wave speed, $\zeta$, for SPS of a rough plate, given different wavelengths ($\lambda$).}
        \label{fig:SPS}
        \end{figure}
        
        We find the Self-Propelled Swimming (SPS) state by setting the wave speed $\zeta$ to zero the mean net thrust $\overline{C_T}$. The wave speed is an effective control parameter to counter roughness adjustments because increasing $\zeta$ increases the thrust production, as shown for a smooth plate in figure \ref{fig:SPS}b. Figure \ref{fig:SPS}a illustrates the change in body shape as the wave speed increases which is a product of fixing the frequency with $St=0.3$ and letting the wavelength change the wave speed. Changing $\zeta$ to achieve SPS allows us to keep $\Rey$, $St$ and $A_1$ constant to test different surface conditions without changing these important swimming parameters identified in the literature. We use Brent's method \citep{Brent1971AnFunction} to find the $\overline{C_T}(\zeta)=0$ root within a tolerance of $10^{-2}$ which allows us to balance precision with the number of iterations; the tabulated solutions are presented in table \ref{tab:forces12}.

        Using this approach, we found that a decrease in the roughness wavelength $\lambda$ requires an increase of $\zeta$ to maintain SPS, figure \ref{fig:SPS}c. $\zeta$ changes with $\lambda$ like the function $\frac{1}{|\lambda|}+c$ (figure \ref{fig:SPS}c). The $\zeta,\lambda$ relationship leads us to restrict $\lambda$ in the range $(1/4, 1/52)$ as the limits are ill-conditioned. Longer wavelengths asymptote to the smooth SPS ($\zeta=1.06$) where $\lambda \equiv 1/0$ whilst all $\lambda<1/52$ are drag-producing for our set-up.

    \subsection{Flow Structures}\label{sec:flow structures}
    
        \begin{figure}
            \centerline{\includegraphics[width=\linewidth]{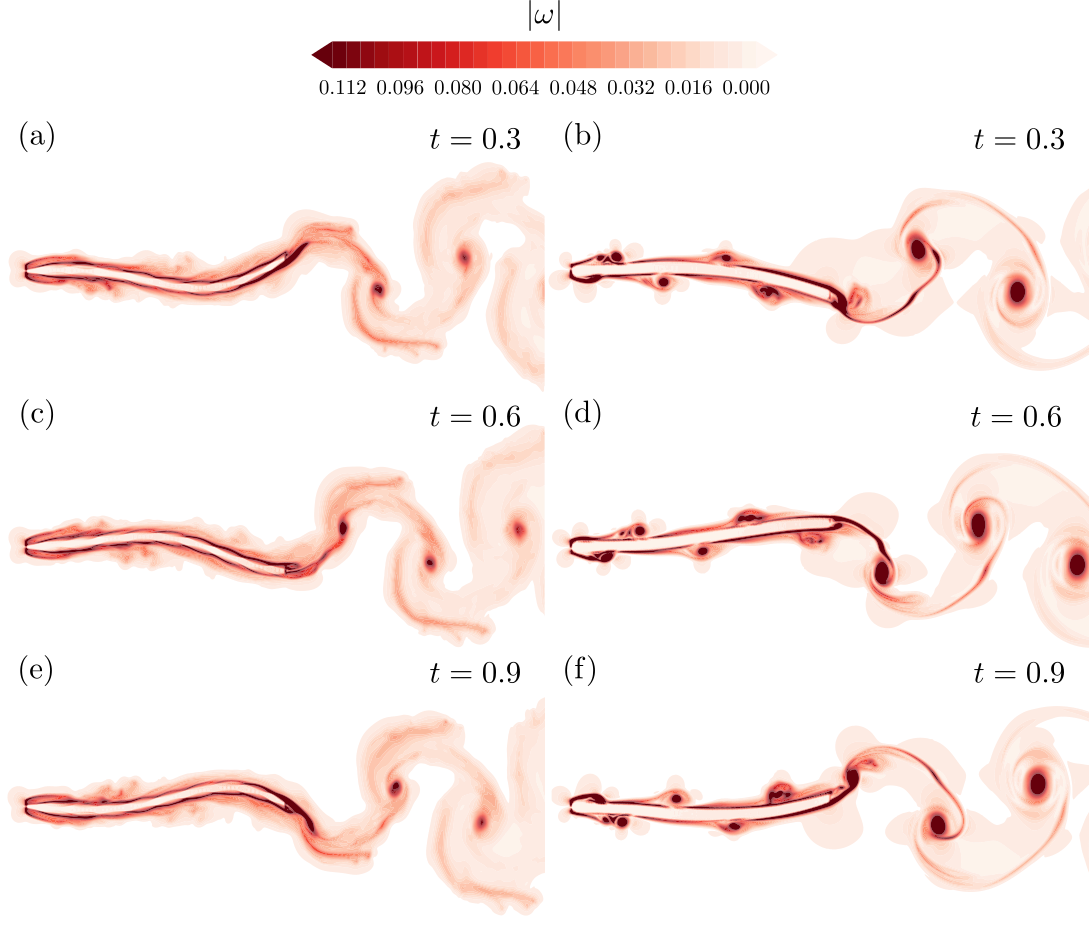}}
            \caption{Sequential snapshots of vorticity magnitude $|\omega|$ for SPS at $\Rey=12,000$. The two columns represent different roughness wavelengths. (a,c,d) $\lambda = 1/4$ requiring $\zeta=1.11$ for SPS. (b,d,e) $\lambda = 1/52$ requiring $\zeta=2.26$.}
            \label{fig:time sequence}
        \end{figure}
    
        \begin{figure}
            \centerline{\includegraphics[width=\linewidth]{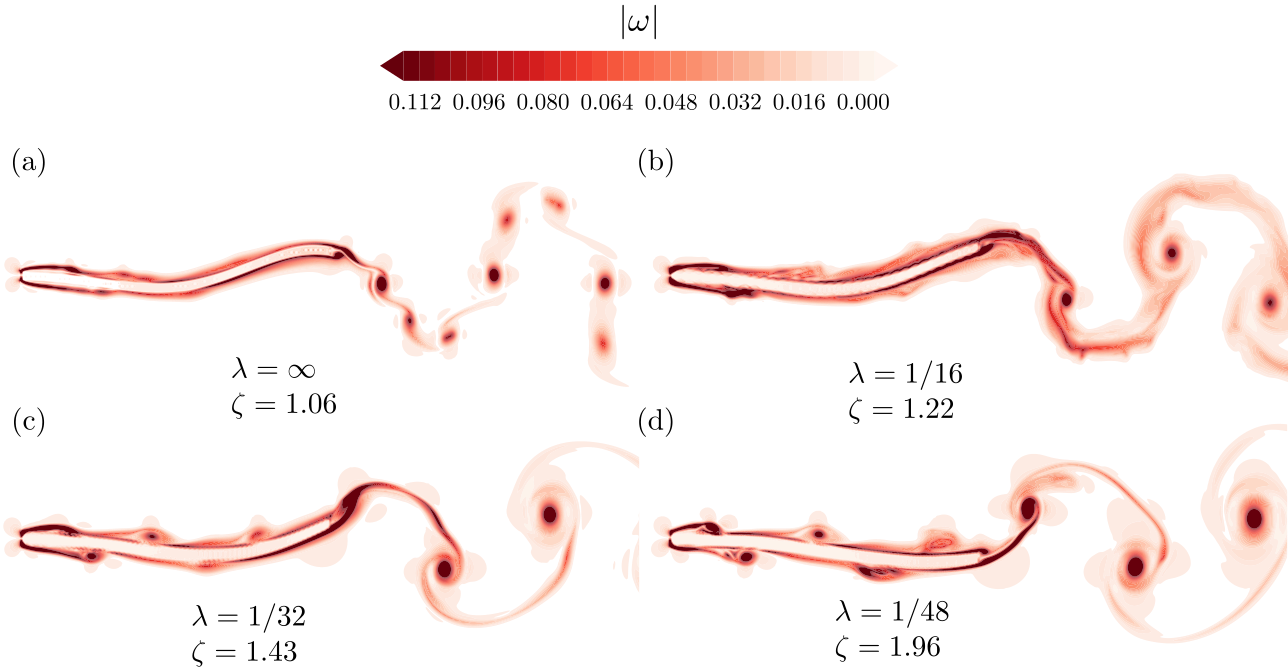}}
            \caption{The change in vorticity magnitude with roughness wavelength. All instances are taken at the same cycle time ($t=0.1$) and show the spanwise-averaged vorticity magnitude. (a) A smooth plate and (b-d) rough plates defined by decreasing $\lambda$.}
            \label{fig:vorticity magnitude}
        \end{figure}

        The structures of the flow provide insight into the workings of the system. 
        Figure \ref{fig:time sequence} shows equally spaced time instances making up a whole period of motion for two different swimming modes at $\Rey=12,000$. One swimming mode is $\zeta=1.11$ which is required to achieve self-propelled swimming for a surface of $\lambda=1/4$, the other is $\zeta=2.27$ for $\lambda=1/52$. The high wave speed required to overcome roughness has resulted in strong coherent vortices whilst the lower wave speed has a much more dispersed vorticity field.
        
        As $\zeta$ increases, the flow around the plate moves away from those typically associated with swimming. Figure \ref{fig:vorticity magnitude} shows four snapshots across the range of $\zeta$ associated with the surfaces tested. (a) is a smooth comparison and exhibits a two-pair plus two-single ($2P+2S$) vortex wake structure \citep{Schnipper2009VortexFoil}. For (b), there is an increase in the boundary layer mixing, and the flow moves back to a more traditional 2P structure. As $\lambda$ decreases further, the leading edge vortex becomes more defined, with (c) and (d) exhibiting well-defined vortices along the length of their body which is generally associated with a heaving instead of a swimming plate
 
\begin{figure}
            \centerline{\includegraphics[width=\linewidth]{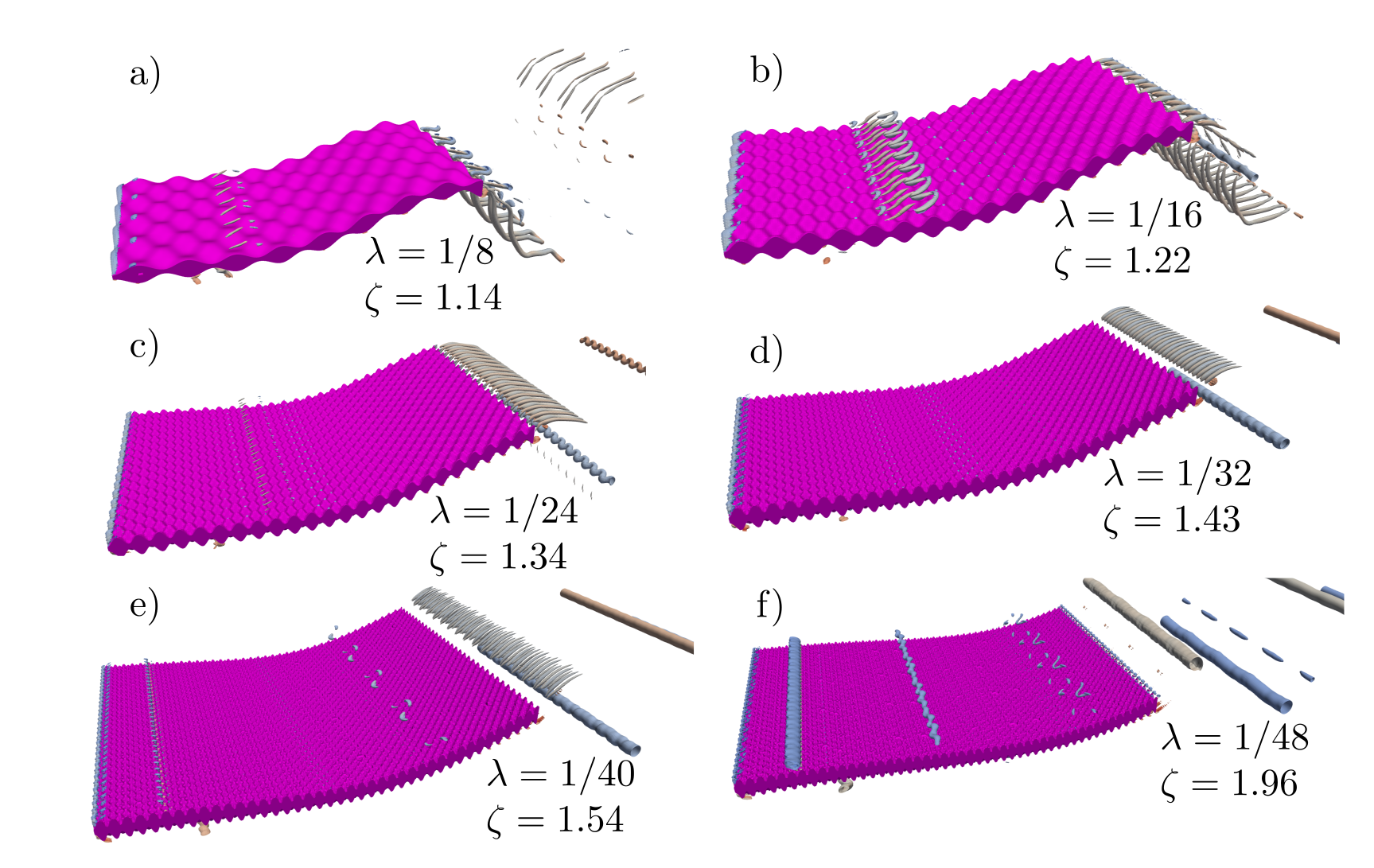}}
            \caption{The Q-Criterion of the flow around a flat plate with a decreasing roughness wavelength. For longer wavelength roughness, the shedding off of the bumps dominate the structures. As the wavelength decreases, the flow transitions to a predominantly two-dimensional state as the influence of the bump perturbations gives way to dominant kinematically-driven structures.}
            \label{fig:q flow structure}
        \end{figure}

        Figure \ref{fig:q flow structure} shows the flow structure visualised by isosurfaces of the Q-Criterion (\citealt{Hunt1988EddiesFlows}). For a direct comparison, the contour of the isosurface remains the same between the figures. The flow exhibits a distinguishable transition as $\zeta$ increases. For low $\zeta$ the bumps dominate the flow structure, and we see distinct horseshoe vortices shed from each element. These vortices persist downstream into counter-rotating streaks that compose the near wake. The flow structures get smaller as $\zeta$ increases and the horseshoe vortex around each element becomes less distinct. From the middle left figure onwards, we can see the near wake collects into a wavy vortex tube similar to those that categorise a two-dimensional flow driven by kinematics \cite{Zurman-Nasution2020}.
        
\begin{figure}
            \centerline{\includegraphics[width=\linewidth]{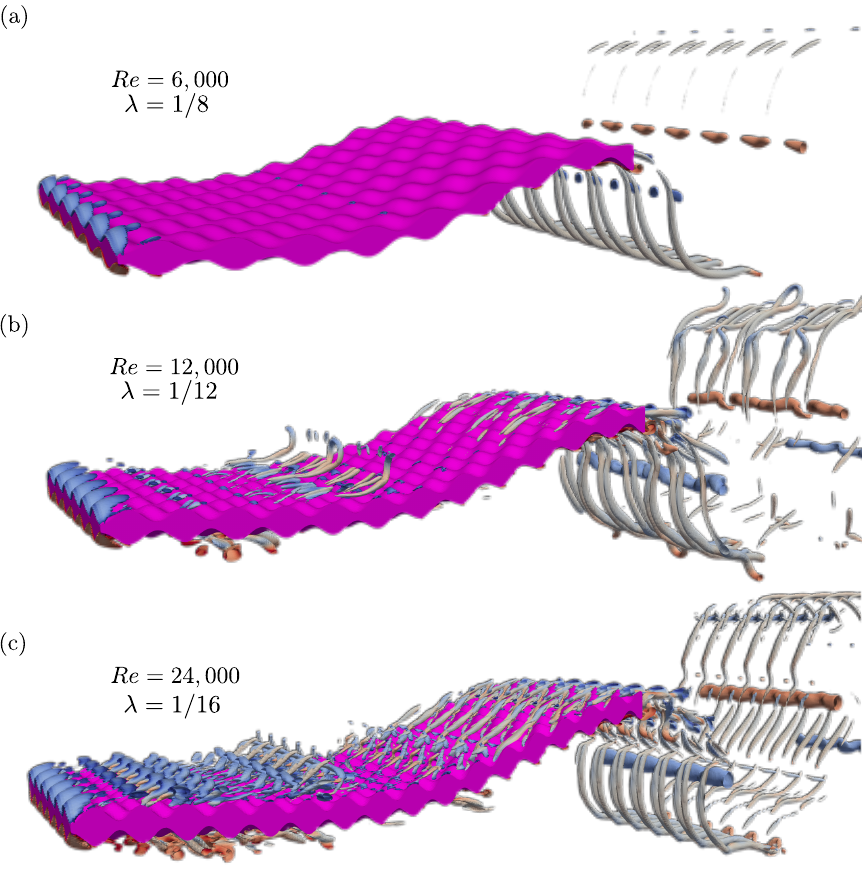}}
            \caption{Flow structures of plates at $\Rey=6,12,$ and $24\times 10^3$ visualised by the same q-criterion. The surface topographies, defined by $\lambda=1/8,1/12,1/16$, correspond to the enstrophy peaks discussed in section \ref{sec:mixing}. An animation of each of these subfigures is available in the supplementary material.}
          \label{fig:o-s-flow}
          \end{figure}

        Next, we study the influence of $\Rey$ on the self-propelled swimming flow by extending the range to $\Rey=6,12$ and $24\times 10^3$, figure \ref{fig:o-s-flow}. The large-scale flow structures such as the leading edge vortex and wake vortices remain similar, but increasing $\Rey$ is seen to greatly increase the production of small-scale vortices - both on the surface and in the wake. This is because the large-scale structures are driven by the kinematics and the surface topology and the reduced viscosity causes the structures to break down quickly.
        
    \subsection{Forces}

\begin{figure}
            \centerline{\includegraphics{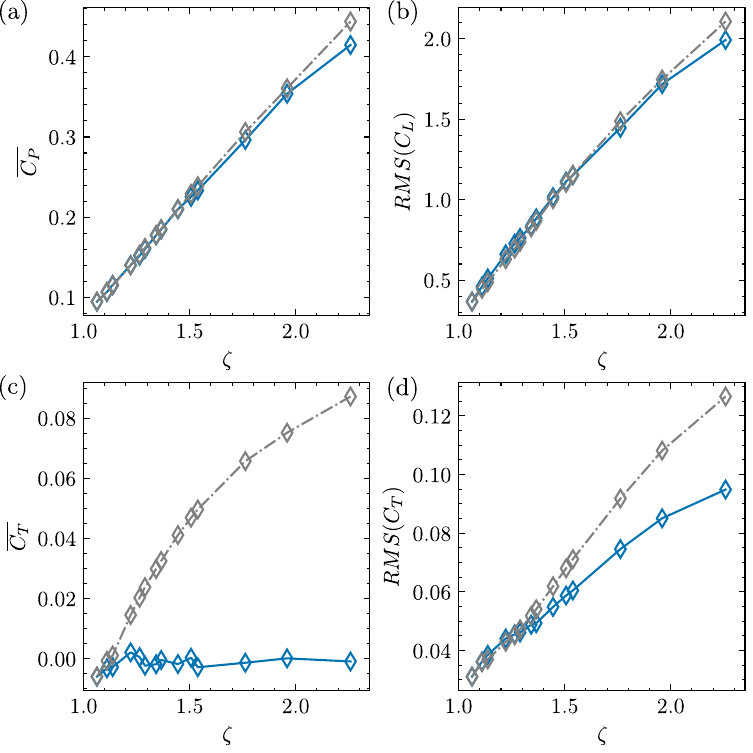}}
            \caption{The key swimming performance and force characteristics of the plate at $\Rey=12,000$. (a) The power required to maintain self-propelled swimming. (b) The side force RMS. (c) The time-averaged and (d) RMS thrust. The points making up the solid line are the rough simulations, and the dash-dot line is a kinematically equivalent smooth simulation which matches $\zeta$ to the rough case.}
            \label{fig:forces}
        \end{figure}

        We find that increasing $\zeta$ increases the power to maintain self-propelled swimming (figure \ref{fig:forces}a) despite the strong two-dimensionality of the flow structures which are normally associated with efficient power transfer \citep{Zurman-Nasution2020} as no energy is lost to three-dimensional effects. This means low $\zeta$ corresponding to the longer wavelength roughness and associated three-dimensional flow structures are more efficient.

        Figure \ref{fig:forces}b show that as $\zeta$ increases, as does $RMS(C_L)$. This signals ineffective swimming as the side forces are balanced by the mass $\times$ acceleration of the body's motion so larger side forces on an equally massive body cause the body to accelerate side to side more, decreasing the smoothness. Similarly, increasing $\zeta$ also increases $RMS(C_T)$ (figure \ref{fig:forces}d) leading to more of a surging motion, further decreasing the smoothness of the swimming. These signs of ineffective swimming are reflected in the stated increase in $\overline{C_P}$.
        
        To decouple the roughness and wave speed effects, we run a smooth simulation with the same kinematics properties as each rough case. This smooth kinematic counterpart gives a base flow to compare against the rough simulations, and the cycled average power and forces are shown in Figure \ref{fig:forces}. Figure \ref{fig:forces}c illustrates that increasing $\zeta$ for the smooth case increases $\overline{C_T}$, making the smooth-plate counterparts slightly thrust producing. The power, $\overline{C_P}$ is reduced by adding roughness to the surface of the plate, Figure \ref{fig:forces}a, for the shorter wavelength roughness tested. Shorter wavelength roughness also reduces $RMS(C_L)$ (figure \ref{fig:forces}b) and $RMS(C_T)$ (figure \ref{fig:forces}d), making the swimming more effective compared to a smooth, kinematic counterpart.


        

    \subsection{Enstrophy}\label{sec:mixing}

        \begin{figure}
          \centerline{\includegraphics{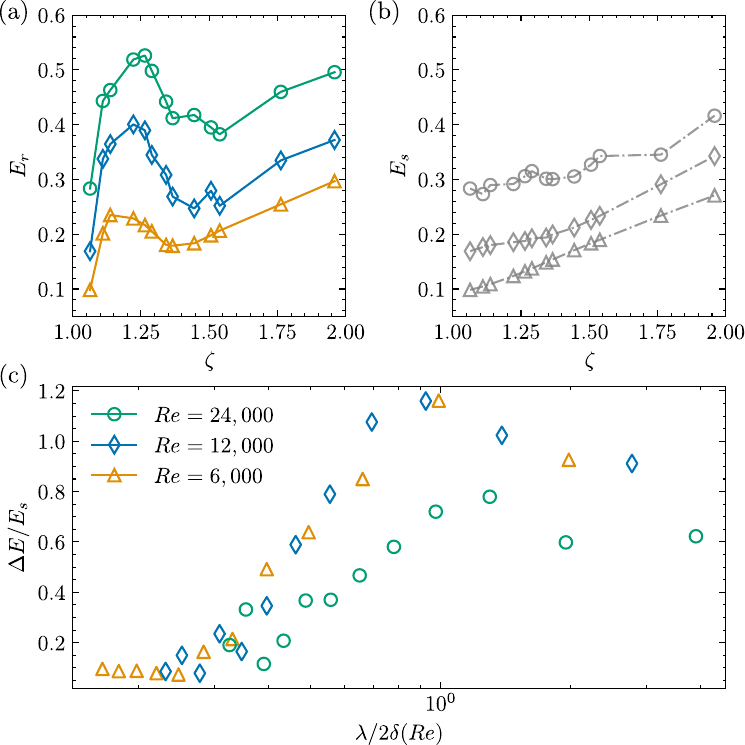}}
          \caption{(a) The $\zeta$-dependant enstrophy for the rough, self-propelled swimming plates ($E_r$) where higher $\zeta$ corresponds to lower $\lambda$. The circle, diamond and triangle markers correspond to $\Rey = 6, 12$ and $24 \times 10^3$ respectively. (b) The enstrophy for the kinematically identical smooth plate $E_s$ where the dash-dot line identifies the data as smooth and the markers relating to the $\Rey$ correspond as before. (c) The difference between the smooth and the rough enstrophies, $\Delta E$ with respect to $\lambda /2\delta(\Rey)$. We take $\delta(\Rey)$ as the approximation of a laminar boundary layer thickness on a smooth, flat plate. The same markers are used to distinguish $\Rey$, and the colours orange, blue and green (online) correspond to $\Rey = 6, 12$ and $24 \times 10^3$ to aid in the distinction.}
        \label{fig:enstrophyscale}
        \end{figure}
    
        So far, we have shown surface roughness increases the drag on a surface, leading to inefficient swimming, and have identified variations in the flow structures for different textures and kinematics. To identify the separate fluid dynamic contribution of $\zeta$ and $\lambda$ we look at scaled enstrophy of the rough surface and its smooth kinematic counterpart. We define the scaled enstrophy as
        
        \begin{equation}
            E = \frac{\int 0.5|\omega|^2 dV }{SA_1}
        \end{equation}\label{eq:enstrophy}
        \\
        where the scaling factor is the planform area times the motion amplitude to define an appropriate reduced volume over which to evaluate the mixing. We use the subscript ($r, s$) to identify the rough and smooth cases respectively. 
        
        In general, the enstrophy increases with $\Rey$ because of the presence of smaller scale structures (figure \ref{fig:o-s-flow}). We show this in figure \ref{fig:enstrophyscale}a,b, which reports the results of $E$ against $\zeta$ for three $\Rey$. Figure \ref{fig:enstrophyscale}a are the results of the rough plate undergoing self-propelled swimming and figure \ref{fig:enstrophyscale}b are the enstrophy for the smooth, kinematic counterparts. Both the smooth and rough plates have an approximately linear increase that scales with $\log{\Rey}$. 
        
        The positive gradient of both the smooth and rough enstrophy shows enstrophy increases with $\zeta$, and the offset between the smooth and the rough line indicates that adding roughness also increases enstrophy. As we increase $\zeta$, the enstrophy increases for both the smooth and the rough cases. At the upper limit of $\zeta$, the smooth and the rough lines start converging as the rough flow, like the smooth, becomes two-dimensional (figure \ref{fig:q flow structure}).

        One prominent feature of figure \ref{fig:enstrophyscale}a are the peaks in enstrophy found between $\zeta=1.1-1.3$. The peak in enstrophy coincides with the superposition of the flow features associated with both the roughness and kinematics (figure \ref{fig:o-s-flow}). This is evident as surfaces with shorter $\lambda$ are increasingly dominated by the two-dimensional vortex tubes associated with the kinematic flow as the structures shed off the roughness elements decrease in size (figure \ref{fig:q flow structure}b-f), and surfaces with longer $\lambda$ are dominated by the flow off the roughness elements (figure \ref{fig:q flow structure}a,b).
        
        The mechanisms driving the peak in enstrophy are analogous to an amplification of flow structures found in Prandtl's Secondary Motions of the Second Kind \citep{JohannNikuradse1926UntersuchungGeschwindigkeitsverteilung, LudwigPrandtl1926UberTurbulenz}; where secondary currents were induced when the surface texture has features that scale with the outer length-scale of the flow. In fact, \citet{Hinze1967SecondaryTurbulence, Hinze1973ExperimentalConduit} found secondary-currents form when surface roughness has dominant scales comparable to the boundary-layer thickness (or pipe/channel height). These secondary currents manifest as low and high-momentum pathways in the flow \citep{Barros2014ObservationsLayer} that are further sustained by spatial gradients \citep{Anderson2015NumericalRoughness}. \citet{Vanderwel2015EffectsLayers} identified that the spatial gradients and the strength of these secondary currents are maximised when the spanwise-spacing between successive roughness features are approximately equal to the boundary layer thickness. When the spacing is small, the flow behaves more like a homogeneous rough surface. When the spacing is much larger than the secondary motions, the currents are spatially confined (and small) to the location of the roughness. The analogy to the case presented in this manuscript is not a direct comparison to Prandtl's Secondary Motions of the Second Kind as the flow is merely unsteady, although, the system studied herein meet certain criteria. Specifically, large-scale streamwise vortical structures (figures \ref{fig:q flow structure} and \ref{fig:o-s-flow}) driven from torque associated with anisotropy of the velocity fluctuations \citep{Perkins1970TheFlow, Bottaro2006FormationFlow}.
        
        The increasing $\Rey$ shifts the peak enstrophy in figure \ref{fig:enstrophyscale}a towards higher $\zeta$ (lower $\lambda$) resulting in flow fields shown in figure \ref{fig:o-s-flow}. The enstrophy peaks occur at $\lambda=1/8, 1/12$ and $1/16$ for $\Rey = 6,12$ and $24 \times 10^3$ respectively. In fact, estimating the value of $\delta (\Rey)$ by assuming a laminar boundary layer correlation ($\delta (\Rey) \approx 4.91{\Rey}^{-0.5}$), and leveraging an approximate scaling $\lambda /2\delta \approx 1$ yields results where $\lambda( \Rey) = 1/7.9, 1/11.5$, and $1/15.8$ for $\Rey =6,12$, and $24 \times 10^3$. These estimations are remarkably close to the true, enstrophy peaking wavelengths, so future studies can use this estimation to assess the importance of their dominant roughness length scales. The laminar boundary layer value of $\delta$ is taken since it is difficult to estimate a value of $\delta$ for the swimming plate. At these Reynolds numbers, $\delta$ could also be estimated using a turbulent boundary layer correlation ($\delta \propto \Rey^{0.2}$) since there is very little difference between the laminar and turbulent values.
        
        
        Further, we can delineate the relationship between $\lambda$ and and the boundary layer thickness by plotting the normalised difference in enstrophy between the smooth and the rough plate ($\Delta E=(E_r-E_s)$) against $\lambda/2\delta(\Rey)$ (figure \ref{fig:enstrophyscale}c). Figure \ref{fig:enstrophyscale}c shows a strong collapse of the $\Delta E/E_s$ curves when plotted against $\lambda/2\delta(\Rey)$ for $\Rey=6$ and $12 \times 10^3$. The scaled maximum of both curves is 1.16 at  $\lambda/2\delta(\Rey)$ just less than 1. The collapse breaks down at the highest $\Rey=24 \times 10^3$ because the flow on the smooth plate experiences a jump in $E_s$ at this Reynolds number, affecting the enstrophy differences.
        
        Finally, in our study, the power increases almost linearly until $\zeta \approx 1.8$ (figure \ref{fig:forces}a) and our enstrophy peak lies within this regime. This means that the accentuation of these secondary flows is truly a boundary layer scaling, and not a result of increased vorticity production at the wall, which would result in an increased force on the body. Therefore, we can relate the system's increase in mixing to scaling arguments previously defined in turbulent wall-flows.

\section{Conclusion} \label{sec:conclusion}

    In this paper, we examined the effect of an egg carton-type rough surface on a Self Propelled Swimming Body. We varied the wavelength of the surface to understand how different surface topologies change the flow and performance of the swimmer. We found a decrease in the roughness wavelength requires a greater wave speed to maintain self-propelled swimming. The greater wave speed changed the vortex structures and consolidated the vorticity into two-dimensional packets with a distinct leading edge vortex propagating down the body. The long wavelength rough surfaces were dominated by the shedding of horseshoe vortices from individual roughness elements that persisted in the wake. The thrust of the plate increased with wave speed, which was needed to overcome the drag induced by the roughness. The increased wave speed also increased the required power and the amplitude of the non-propulsive lift force. These increases implied the plate was less efficient, less effective, and less steady in its swimming. To decouple the effects of roughness and kinematics we compared the forces and enstrophy to a smooth swimmer with identical kinematics. We saw that, compared to the smooth cases, the roughness reduced the power required, as well as the amplitude of the lift and drag forces.
    There was a peak in enstrophy which coincided with a superposition of two flow modes, one dominated by three-dimensional structures and the other by the two-dimensional vortex tubes. The peak in enstrophy persisted over all three Reynolds numbers, and collapsed when the roughness wavelength is proportional to the boundary layer thickness. This spike in enstrophy is not a result of increased vorticity production at the wall because we do not have a corresponding increase in body force. Further, this boundary-layer relationship is analogous to scaling arguments defined for turbulent wall-flows (\citealt{Vanderwel2015EffectsLayers}).
    
    These results show that you cannot ignore kinematics when assessing the performance of a swimmer with surface texture. \citet{Oeffner2012} and \citet{Wen2014} reported an increase in speed and efficiency for their experiments, but we have shown a change in wave speed dominates the thrust production of the plate. Adding a coating to a surface could increase the stiffness and thus increase the wave speed, causing the plate to swim faster. Another factor to consider is the structural resonance, which significantly affects the performance of flexible plates undergoing swimming (\citealt{Quinn2014ScalingPanels}). Any study attributing performance changes to surface textures must show the independence of their test cases to the kinematics.

    We have identified non-linear interactions between the roughness and kinematics that amplify this mixing without a nonlinear force or power increase. Other studies (\citealt{lang2008bristled, afroz2016experimental, Santos2021}) have identified the bristling of shark skin and conjectured that the increased mixing helps keep flow attached in the flank region. This work is significant in understanding the hydrodynamic effect of surface textures on the flow and forces around a swimmer, it is the first study to look at surface textures on undulating surfaces with realistic and well-defined kinematics. However, it is limited in that the roughness elements are a hundred times larger than actual denticles, and $\Rey$ is only representative of a small, slow-swimming shark. 
    
    
\section*{Acknowledgements}
    We would like to thank the IRIDIS high performance computing facility, and it's associated support at the University of Southampton, the Office of Naval Research Global Award N62909-18-1-2091.

\section*{Declaration of Interests}
    The authors report no conflict of interest.

\section*{Data statement}
    All data for this manuscript will be made available upon publication.
    
\FloatBarrier

\bibliographystyle{jfm}
\bibliography{references}

\appendix
\FloatBarrier
\section{Resolution convergence}\label{sec:convergence}

\begin{figure}
        \centerline{\includegraphics{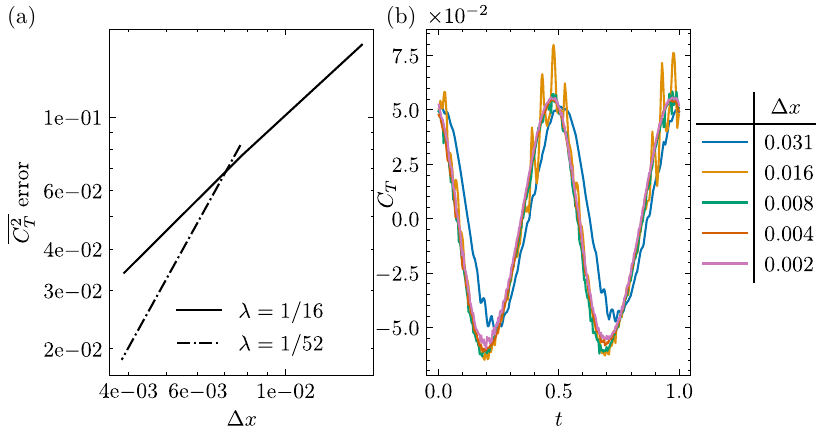}}
        \caption{This figure shows the resolution convergence for the rough, self-propelled swimming plate. (a) The error convergence of $\overline{C_T^2}$. (b) Phase averaged cycle where $\lambda=1/16$.}
    \label{fig:pressure convergence}
    \end{figure}

    We tested resolutions of increasing powers of two for two surfaces where $\lambda=1/16,1/52$ (figure \ref{fig:pressure convergence}). For figure \ref{fig:pressure convergence}a we measured the error against the value at the highest resolution, which contained $\num{2.4e9}$ grid cells. The pressure-based thrust $\overline{C_{T}}$ oscillates around a zero mean, and so we measure the error in $\overline{C_{T}^2}$. We converge to below $4\%$ error for both surfaces. Our working resolution is at the lowest limit of $\Delta x=0.004$ and figure \ref{fig:pressure convergence}b shows that the time history of $C_T$ also converges within this.

\begin{figure}
        \centerline{\includegraphics{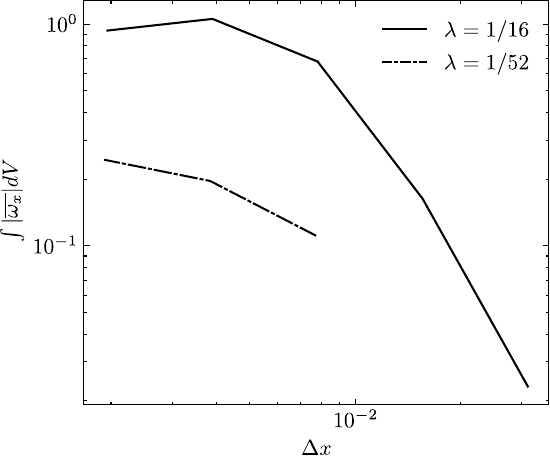}}
        \caption{The convergence of $\int |\overline{\omega_x}| dV $ with simulation fidelity.}
    \label{fig:bump convergence}
    \end{figure}

    Figure \ref{fig:bump convergence} shows the convergence of the integral quantity of the x-vorticity magnitude for the two surfaces where $\lambda=1/16,1/52$.  We choose the surface where $\lambda=1/16$ because it is around this value that significant increases in enstrophy were found. We also test the surface where $\lambda=1/52$ because this represents the lowest limit of our surface resolution; for the proposed working resolution of $\Delta x=0.004$, $5$ grid cells resolve the surface wavelength in the $x$ and in the $z$ direction.
    Furthermore, we measure $\int |\omega_x| dV$ because it is zero for the two-dimensional, smooth cases and, therefore, allows us to quantify the grid-resolution-independence of the topographic contribution to the flow. Again, we converged to within a reasonable limit at our working resolution of $\Delta x=0.004$.

\section{Domain study}\label{sec:domain}

\begin{figure}
        \centerline{\includegraphics{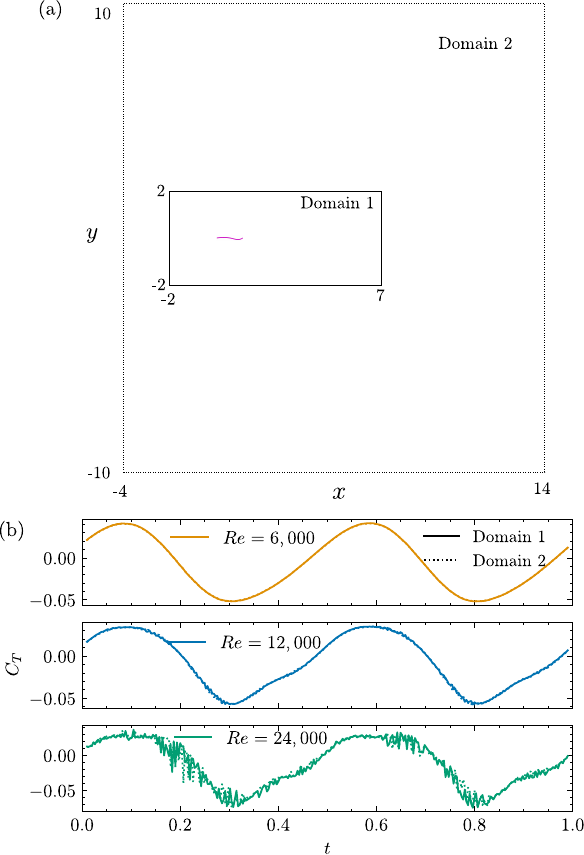}}
        \caption{The domain invariance of $C_T$. (a) The relative size of the two-dimensional domains. The dotted line marks the extent of Domain 2 where $(x, y) \in ([-4, 14], [-10, 10])$, and the solid line marks Domain 1, where $(x, y) \in ([-2, 7], [-2, 2])$. (b) The thrust coefficient for $\zeta=1.06$ and $St=0.3$, the solid line for Domain 1 and the dotted line of Domain 2 are plotted on top of each other.}
        \label{fig:domain}
    \end{figure}

    We show the domain invariance of the solution by comparing the cycle average time series of $C_T$. We compare the working domain (Domain 1) of size $(18,20)$ and resolution $(1536,1536)$, to a much larger domain (Domain 2), size $(9,4)$ and resolution $(3072,3072)$ (figure \ref{fig:domain}a). Figure \ref{fig:domain}b shows the cycle average time series of $C_T$ for the three $\Rey$ used in this study. For all three ($\Rey=6,12,$ and $24\times 10^3$), $C_T$ is plotted for Domain 1 (solid line), and Domain 2 (dotted line). $C_T$ for both domains collapses, and so the solution is domain invariant.
    
    \section{Experimental validation}\label{sec:kin-validation}
    
\begin{figure}
        \centerline{\includegraphics{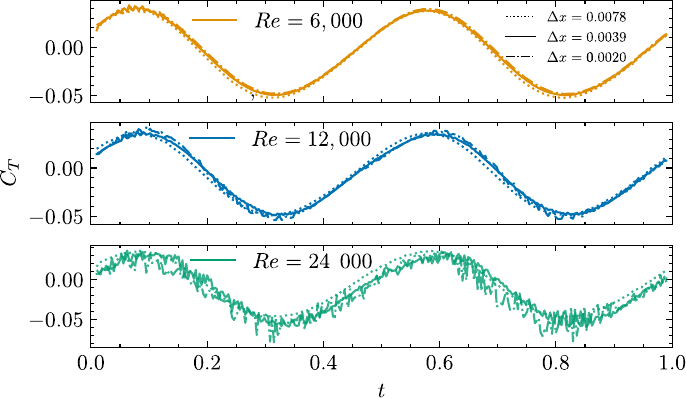}}
        \caption{The numerical convergence for the kinematic trajectory defined by equation \ref{eq:kinematic trajectory} for $\zeta=1.06$ and $St=0.3$.}
        \label{fig:kin-convergence}
    \end{figure}

    First, we show that the method implemented in this study, set out in section \ref{sec:Kinematics}, converges to grid-independent solution. We perform simulations at $St=0.3$, $\zeta=1.06$ and $\Rey = 6,12,$ and $24\times 10^3$ with increasing resolution. In figure \ref{fig:kin-convergence}, we plot $C_T$ (phase averaged over four cycles) and show that the resolution minimally affects the time series to the point where different resolutions are barely distinguishable from each other.
            
    
    Next, we validate our model by showing consistency with an experimental study \citep{Lucas2015EffectsModel}. We make minor alterations to our model to match the kinematic trajectory, $St$ and $\Rey$ of a swimming plate. In \citet{Lucas2015EffectsModel}, they test four different plate stiffnesses and assess the swimming performance. Of these four plates one (plate `$1\_3$') had an increasing amplitude envelope and a relatively constant wave speed; which ensures the changes we need to match their conditions are minimal. Although \citet{Lucas2015EffectsModel} test a range of different $St$s and $\Rey$; for the plate we are matching, they only provide the kinematic trajectory for $\Rey=77,000$ and $St=0.31$. Because of the sensitivity of the kinematic mode shape of a flexible plate to different excitation conditions \citep{Quinn2014ScalingPanels}, we can only consider the data point where $\Rey=77,000$ and $St=0.31$.
    
\begin{figure}
        \centerline{\includegraphics{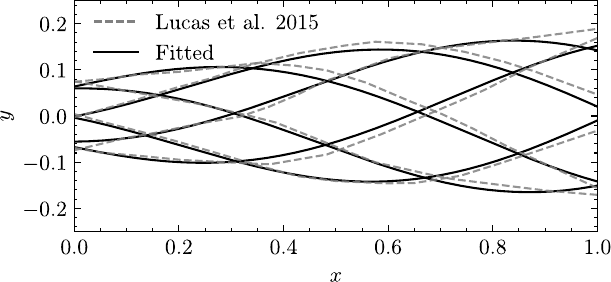}}
        \caption{The difference between the kinematic trajectories reported in \citet{Lucas2015EffectsModel} and the fitted functional form (equation \ref{eq:fitted-lucas}).}
        \label{fig:lucas-trajectory}
    \end{figure}

    Figure \ref{fig:lucas-trajectory} shows a comparison of the raw and matched kinematic trajectories. We match the trajectory by minimising $||y(x,t)-\langle y(x,t) \rangle||^2_2$ where $\langle y(x,t) \rangle$ is the model for the kinematics. We consider the amplitude envelope, $A(x)$ and the wave speed, $\zeta$ separately and arrive at the functional form
    
    \begin{equation}
        \langle y(x,t) \rangle = a_i x^i \sin{\big(2\pi(x/\zeta - f t)\big)}
    \end{equation}\label{eq:fitted-lucas}
    \\
    where $i_{0,1,2}=[0.072, 0.1685, -0.0701]$, and $\zeta = 2.15$. Furthermore, we measured the error in $\langle y(x,t) \rangle$ using

    \begin{equation}
        \sigma(y(x,t)-\langle y(x,t) \rangle)/\sigma(y(x,t))=0.076
    \end{equation}
    \\
    where $\sigma(x) = \sqrt{\overline{(x-\overline{x})^2}}$ is the operator to compute the standard deviation. The raw data in \citet{Lucas2015EffectsModel} does not perfectly match the idealised model of a swimmer, with slight asymmetry (figure \ref{fig:lucas-trajectory}) introducing the reported error.
    
    We have kept the domain size, grid-refinement and time-step constant to our working model and only altered the kinematic trajectory, $St$ and $\Rey$. This allows us to assess the accuracy of the computational setup compared to \citet{Lucas2015EffectsModel}. As we increase the resolution, $\overline{C_T}$ converges to within $1.7\%$ of the reported experimental value, table \ref{tab:converge}. This gives us confidence that the computational setup is accurate and that the kinematics are correctly implemented. 
    
    \begin{table}
        \begin{center}
            \def~{\hphantom{0}}
            \begin{tabular}{ccc}
                $\Delta x$ & $\overline{C_T}$ & error \\
                $0.00391$ & $0.183$ & $0.131$ \\
                $0.00195$ & $0.167$ & $0.029$ \\
                $0.00098$ & $0.165$ & $0.017$
            \end{tabular}
        \end{center}
        \caption{The convergence of $\overline{C_T}$ to experimental results of \citet{Lucas2015EffectsModel}.}
        \label{tab:converge}
    \end{table}

    \section{Tabulated forces}

    Table \ref{tab:forces6}, \ref{tab:forces12}, and \ref{tab:forces24} show $\zeta, \overline{C_T}, \overline{C_P }$ for the rough, and kinematically equivalent smooth simulations. Figure \ref{fig:forces} is the graphical representation but, for ease of comparison, we have also tabulated the data.

    \begin{table}
        \begin{center}
            \def~{\hphantom{-}}
            \begin{tabular}{cccccc}
                $\lambda$ & $\zeta$ & $\overline{C_T}$& $\overline{C_T}_s$ &  $\overline{C_P}$&  $\overline{C_P}_s$ \\
                1/0~ & $1.06$ & $-0.007$ & $-0.007$ & $0.088$ & $0.088$ \\
                1/4~ & $1.11$ & $-0.003$ & $-0.000$ & $0.103$ & $0.102$ \\
                1/8~ & $1.14$ & $-0.003$ & $~0.004$ & $0.111$ & $0.110$ \\
                1/12 & $1.22$ & $~0.002$ & $~0.015$ & $0.135$ & $0.135$ \\
                1/16 & $1.26$ & $~0.000$ & $~0.020$ & $0.148$ & $0.148$ \\
                1/20 & $1.29$ & $-0.003$ & $~0.023$ & $0.156$ & $0.156$ \\
                1/24 & $1.34$ & $-0.002$ & $~0.029$ & $0.172$ & $0.173$ \\ 
                1/28 & $1.37$ & $-0.002$ & $~0.031$ & $0.179$ & $0.180$ \\
                1/32 & $1.45$ & $-0.003$ & $~0.039$ & $0.204$ & $0.205$ \\
                1/36 & $1.51$ & $-0.000$ & $~0.045$ & $0.221$ & $0.224$ \\
                1/40 & $1.54$ & $-0.004$ & $~0.048$ & $0.229$ & $0.234$ \\
                1/44 & $1.76$ & $-0.001$ & $~0.064$ & $0.287$ & $0.299$ \\
                1/48 & $1.96$ & $~0.000$ & $~0.074$ & $0.340$ & $0.355$ \\
                1/52 & $2.26$ & $-0.003$ & $~0.084$ & $0.396$ & $0.417$ \\                    
            \end{tabular}
        \end{center}
        \caption{Tabulated data of simulations at $\Rey=6,000$ with the input roughness defined by $\lambda$ and corresponding $\zeta$ that results in self-propelled swimming. The table also reports the values for $\overline{C_T}, \overline{C_P}$ where the subscript $(\cdot)_s$ refers to a smooth plate for comparison.}
    \label{tab:forces6}
    \end{table}

    \begin{table}
        \begin{center}
            \def~{\hphantom{-}}
            \begin{tabular}{cccccc}
                $\lambda$ & $\zeta$ & $\overline{C_T}$& $\overline{C_T}_s$ &  $\overline{C_P}$&  $\overline{C_P}_s$ \\
                1/0~ & $1.06$ & $-0.006$ & $-0.006$ & $0.095$ & $0.095$ \\
                1/4~ & $1.11$ & $-0.003$ & $-0.001$ & $0.107$ & $0.107$ \\
                1/8~ & $1.14$ & $-0.003$ & $~0.001$ & $0.116$ & $0.115$ \\
                1/12 & $1.22$ & $~0.002$ & $~0.015$ & $0.140$ & $0.140$ \\
                1/16 & $1.26$ & $~0.001$ & $~0.020$ & $0.152$ & $0.153$ \\
                1/20 & $1.29$ & $-0.002$ & $~0.024$ & $0.160$ & $0.162$ \\
                1/24 & $1.34$ & $-0.002$ & $~0.030$ & $0.177$ & $0.178$ \\ 
                1/28 & $1.37$ & $-0.000$ & $~0.033$ & $0.185$ & $0.185$ \\
                1/32 & $1.45$ & $-0.002$ & $~0.041$ & $0.210$ & $0.210$ \\
                1/36 & $1.51$ & $~0.000$ & $~0.047$ & $0.225$ & $0.229$ \\
                1/40 & $1.54$ & $-0.003$ & $~0.050$ & $0.233$ & $0.239$ \\
                1/44 & $1.76$ & $-0.001$ & $~0.066$ & $0.296$ & $0.306$ \\
                1/48 & $1.96$ & $~0.000$ & $~0.075$ & $0.354$ & $0.361$ \\
                1/52 & $2.26$ & $-0.001$ & $~0.087$ & $0.415$ & $0.444$ \\                    
            \end{tabular}
        \end{center}
        \caption{Tabulated data of simulations at $\Rey=12,000$ with the input roughness defined by $\lambda$ and corresponding $\zeta$ that results in self-propelled swimming. The table also reports the values for $\overline{C_T}, \overline{C_P}$ where the subscript $(\cdot)_s$ refers to a smooth plate for comparison.}
    \label{tab:forces12}
    \end{table}

    \begin{table}
        \begin{center}
            \def~{\hphantom{-}}
            \begin{tabular}{cccccc}
                $\lambda$ & $\zeta$ & $\overline{C_T}$& $\overline{C_T}_s$ &  $\overline{C_P}$&  $\overline{C_P}_s$ \\
                1/0~ & $1.06$ & $-0.011$ & $-0.011$ & $0.097$ & $0.097$ \\
                1/4~ & $1.11$ & $-0.003$ & $-0.001$ & $0.109$ & $0.111$ \\
                1/8~ & $1.14$ & $-0.002$ & $~0.000$ & $0.117$ & $0.119$ \\
                1/12 & $1.22$ & $~0.002$ & $~0.015$ & $0.141$ & $0.147$ \\
                1/16 & $1.26$ & $~0.000$ & $~0.019$ & $0.153$ & $0.161$ \\
                1/20 & $1.29$ & $-0.004$ & $~0.021$ & $0.161$ & $0.169$ \\
                1/24 & $1.34$ & $-0.004$ & $~0.026$ & $0.176$ & $0.183$ \\ 
                1/28 & $1.37$ & $-0.003$ & $~0.030$ & $0.184$ & $0.191$ \\
                1/32 & $1.45$ & $-0.005$ & $~0.040$ & $0.212$ & $0.217$ \\
                1/36 & $1.51$ & $-0.000$ & $~0.046$ & $0.229$ & $0.236$ \\
                1/40 & $1.54$ & $-0.004$ & $~0.049$ & $0.238$ & $0.246$ \\
                1/44 & $1.76$ & $-0.001$ & $~0.067$ & $0.302$ & $0.310$ \\
                1/48 & $1.96$ & $~0.002$ & $~0.077$ & $0.361$ & $0.365$ \\
                1/52 & $2.26$ & $~0.001$ & $~0.088$ & $0.420$ & $0.451$ \\                    
            \end{tabular}
        \end{center}
        \caption{Tabulated data of simulations at $\Rey=24,000$ with the input roughness defined by $\lambda$ and corresponding $\zeta$ that results in self-propelled swimming. The table also reports the values for $\overline{C_T}, \overline{C_P}$ where the subscript $(\cdot)_s$ refers to a smooth plate for comparison.}
    \label{tab:forces24}
    \end{table}
    
    \end{document}